\documentclass{llncs}
\usepackage{amsmath}
\usepackage{amssymb}
\usepackage{enumerate}
\usepackage{amsfonts}
\usepackage{graphicx}
\usepackage{xspace}
\usepackage{mathrsfs}

\usepackage{verbatim}

\usepackage{url}

\newcommand{\lodz}{{\L}\'{o}d\'{z}}

\newcounter{instr}
\newcommand{\ninstr}{\refstepcounter{instr}\theinstr.}
\newcommand{\bigO}{\mathcal{O}\xspace}

\newcommand{\msampling}{\textsc{M-Sampling}\xspace}

\newcommand{\slength}[1]{len(#1)}

\newcommand{\defAs}
{\mathrel{\makebox{\:= \hspace{-.2cm} \raisebox{-0.5 ex}[0 ex][0 ex]{\tiny Def}}}}

\newcommand{\mdsuff}{\sqsupseteq_{md}}

\newlength{\gnat}

\newlength{\gnatb}

\newcommand{\vect}{G}
\newcommand{\fgg}{\textsc{GFG}\xspace}
\newcommand{\fgga}{\textsf{GFG1}\xspace}
\newcommand{\fggb}{\textsf{GFG2}\xspace}
\newcommand{\ms}{\textsf{MS}\xspace}

\newcommand{\abs}[1]{\emph{abs}(#1)}
\newcommand{\Abs}[1]{\emph{abs}\big(#1\big)}
\newcommand{\sht}{|}
\newcommand{\occ}{occ}

\title{String Matching\\with Inversions and Translocations\\in Linear Average Time (Most of the Time)}
\author{Szymon Grabowski$^\dag$ \and Simone Faro$^\ddag$ \and Emanuele Giaquinta$^\ddag$}
\institute{$^\dag$ Computer Engineering Department, Technical University of \lodz,\\Al. Politechniki 11, 90-924 \lodz, Poland\\
	  \email{sgrabow@kis.p.lodz.pl}\\[0.2cm]
	  $^\ddag$ Universit\`a di Catania, Dipartimento di Matematica e Informatica\\Viale Andrea Doria 6, I-95125 Catania, Italy\\
	  \email{\{faro $\mid$ giaquinta\}@dmi.unict.it}
}

\begin{document}

\maketitle

\begin{abstract}
We present an efficient algorithm for finding all approximate occurrences 
of a given pattern $p$ of length $m$ in a text $t$ of length $n$ 
allowing for translocations of equal length adjacent factors and
inversions of factors.
The algorithm is based on an efficient filtering method and has an
$\bigO(nm\max(\alpha, \beta))$-time complexity in the worst case
and $\bigO(\max(\alpha, \beta))$-space complexity,
where $\alpha$ and $\beta$ are respectively the maximum length of the 
factors involved in any translocation and inversion.
Moreover we show that under the assumptions of equiprobability and 
independence of characters our algorithm has a 
$\bigO(n)$ average time complexity, whenever 
$\sigma = \Omega(\log m / \log\log^{1-\varepsilon} m)$, 
where $\varepsilon > 0$ and $\sigma$ is the dimension of the alphabet.
Experiments show that the new proposed algorithm
achieves very good results in practical cases.
\end{abstract}

\section{Introduction}\label{sec:Introduction}
Retrieving information and teasing out the meaning of biological
sequences are central problems in modern biology.  Generally, basic
biological information is stored in strings of nucleic acids (DNA,
RNA) or amino acids (proteins).  Aligning sequences helps in revealing
their shared characteristics, while matching sequences can infer
useful information from them.
With the availability of large amounts of DNA data, matching of
nucleotide sequences has become an important application and there is
an increasing demand for fast computer methods for analysis and data
retrieval.

\emph{Approximate string matching} is a fundamental problem in text
processing and consists in finding approximate matches of a pattern in
a string.  The closeness of a match is measured in terms of the sum of
the costs of the edit operations necessary to convert the string
into an exact match.
Most classical models, e.g., Levenshtein or Damerau edit distance
(for a survey see~\cite{Nav01}) assume that changes between strings
occur locally.
However, evidence shows that large scale changes are possible in 
chromosomal rearrangment.
For example, large pieces of DNA in a chromosomal sequence can be broken 
and moved from one location to another.
This is known as a \emph{chromosomal translocation}.
Sometimes a mutation can also flip a stretch of DNA within a chromosome,
producing a \emph{chromosomal inversion}.

\begin{figure}[!t]
\begin{tabular}{cc}
\includegraphics[width=0.48\textwidth,trim=0 0 250 0,clip=true]{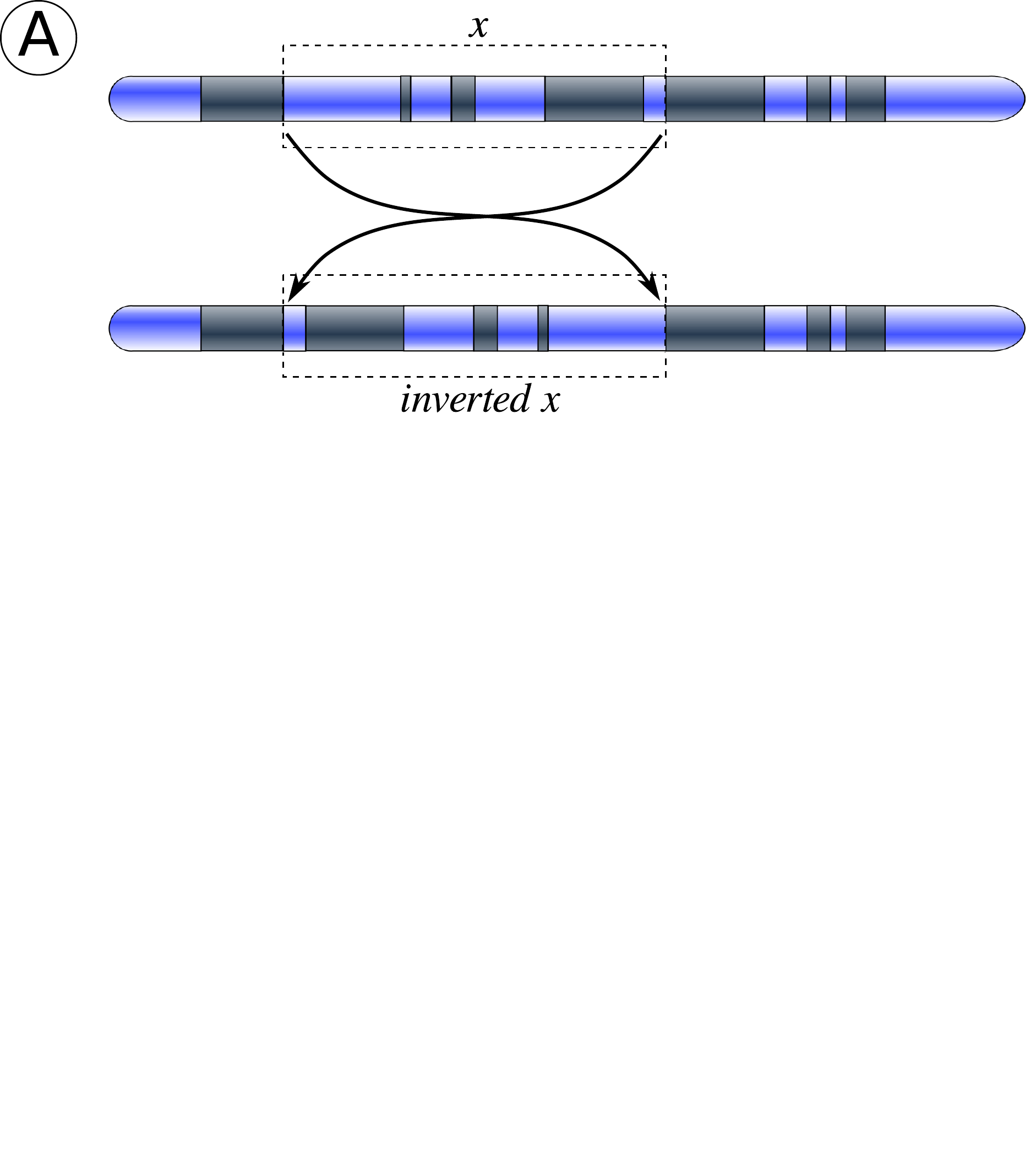} &
\includegraphics[width=0.48\textwidth,trim=0 0 250
-100,clip=true]{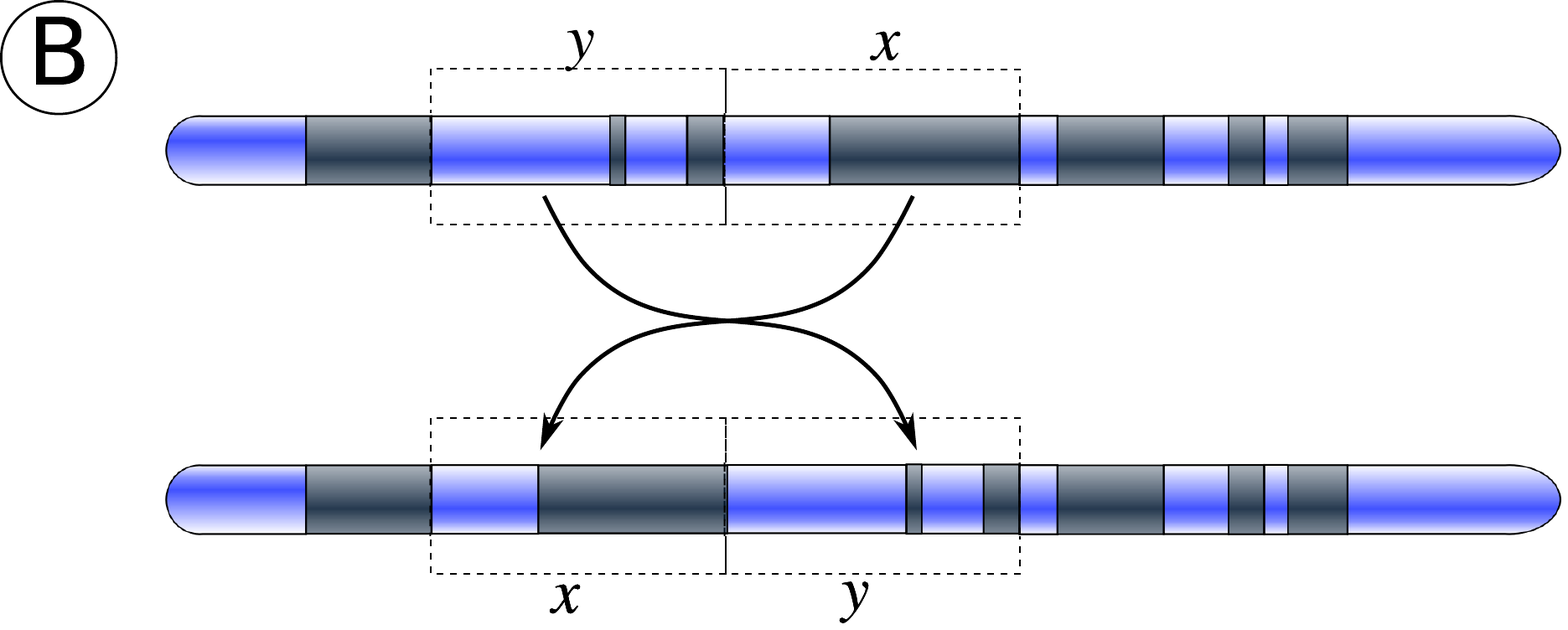}\vspace{-4.5cm}\\
\end{tabular}
\caption{An example of chromosomal inversion.}
\label{fig:mutations}
\end{figure}

In particular a chromosomal inversion is a rearrangement in which a segment 
of a chromosome is reversed end to end.
An inversion occurs when a single chromosome undergoes breakage and 
rearrangement within itself.
Fig. \ref{fig:mutations}(A) shows an example of chromosomal inversion.

Differently a chromosomal translocation is a chromosome abnormality caused by rearrangement
of parts of the same chromosome or between nonhomologous chromosomes.
Sometimes a chromosomal translocation could join two separated genes, the occurrence of which is common in cancer.
Fig. \ref{fig:mutations}(B) shows an example of chromosomal translocation.

Both inversions and translocations do not involve a loss of genetic information, 
but simply rearrange the linear gene sequence.
\smallskip

Recently Cantone et al.~\cite{PSC2010-4} presented the first solution for
the matching problem under a string distance whose edit operations are translocations of
equal length adjacent factors and inversions of factors.  In
particular, they devised a $\bigO(nm\max(\alpha,\beta))$-time and
$\bigO(m^2)$-space algorithm, where $\alpha$ and $\beta$ are the
maximum length of the factors involved in a translocation and in an
inversion, respectively.  They showed that under the
assumption of equiprobability and independence of characters in the
alphabet,
on average the algorithm has a $\bigO(n\log_{\sigma} m)$-time complexity.
Moreover they also presented
a bit-parallel implementation of their algorithm,
which has
$\bigO(n\max(\alpha,\beta))$-time and $\bigO(\sigma + m)$-space
complexity, if the pattern length is comparable with the computer word size.

In this paper we present a new algorithm for the same problem based on
an efficient permutation filtering method and on a dynamic programming approach
for testing candidate positions.
In particular our algorithm
achieves a $\bigO(nm\max(\alpha, \beta))$-worst case time complexity, as the \msampling algorithm, and requires
only $\bigO(\max(\alpha, \beta))$ space. Moreover we show that under the
assumption of equiprobability and independence of characters in the
alphabet, on the average our filter based algorithm achieves a $\bigO(n)$-time complexity,
when $\sigma = \Omega(\log m / \log\log^{1-\varepsilon} m)$, where $\varepsilon > 0$ and $\sigma$ is the dimension of the alphabet.

A slightly shorter version of this manuscript was submitted to 
{\em Information Processing Letters}.

\section{Basic notions and definitions}\label{sec:Notions}
Let $p$ be a string of length $m \geq 0$, over 
an integer alphabet $\Sigma$ of size $\sigma$.
We represent it as a finite array $p[0\, \ldots \,m-1]$ of characters 
from $\Sigma$ and write $\slength{p} = m$.
In particular, for $m=0$ we obtain the empty string $\varepsilon$.
We denote by
$p[i]$ the $(i+1)$th character of $p$, for $0\leq i< m$.  Likewise,
the substring (also called \emph{factor}) of $p$ contained between the $(i+1)$th and the
$(j+1)$th characters of $p$ is indicated with $p[i\, \ldots \,j]$, for $0\leq i
\leq j < m$. An $m$-substring (or $m$-factor) is a substring of length $m$.
We also put $p_i \defAs p[0\, \ldots \,i]$, for $0\leq i<m$.
In addition, we
write $pp'$ to denote the concatenation of $p$ and $p'$, and $p^{\mathsf{r}}$ for
the reverse of the string $p$, i.e., $p^{\mathsf{r}} \defAs p[m-1]p[m-2]\ldots p[0]$. 
Given a string $p$ and a character $c\in \Sigma$,
we define $\occ_p(c)$ as the number of times the character $c$ occurs in $p$ 
(observe that $0\leq \occ_p(c) \leq \slength{p}$).

A \emph{distance} $d:\Sigma^{*}\times \Sigma^{*}\rightarrow
\mathbb{R}$ is a function which associates to any pair of strings $X$
and $Y$ the minimal cost of any finite sequence of edit operations
which transforms $X$ into $Y$, if such a sequence exists, $\infty$
otherwise.

\begin{definition}
Given two strings $X$ and $Y$, the \emph{mutation distance} $md(X,Y)$
is based on the following edit operations:
\begin{itemize}
\item[(1)] \textbf{Translocation}: a factor of the form $ZW$ is
transformed into $WZ$, provided that $\slength{Z}=\slength{W} > 0$.

\item[(2)] \textbf{Inversion}: a factor $Z$ is tranformed into $Z^{\mathsf{r}}$.
\end{itemize}
Both operations are assigned unit cost.\qed
\end{definition}

We indicate with $\alpha$ and $\beta$ the maximum
length of factors involved translocations and inversions, respectively.
By definition, $\alpha \leq \lfloor \slength{X}/2\rfloor$ 
and $\beta \leq \slength{X}$.
When $md(X,Y) < \infty$, we say that $X$ and $Y$ have an $md$-match. 
Additionally, if $X$ has an $md$-match with a suffix of $Y$, we write $X \mdsuff Y$.

\section{Proposed Algorithm}\label{sec:new}
In this section we present a new efficient algorithm for the 
approximate string matching problem allowing for
inversions of factors and translocations of equal length adjacent factors. 
In the following we assume that $p$ and $t$ are strings
of length $m$ and $n$ respectively, over a common alphabet 
$\Sigma =\{c_0, \ldots, c_{\sigma-1}\}$, where $\sigma = O(n)$.
(The case of even larger alphabets is rather theoretical and can be handled 
with standard solutions, e.g., using a minimal perfect hash function.)

The new algorithm, named \fgg algorithm, searches for all occurrences 
of $p$ in $t$ by making use of an efficient filter method.
This technique, usually called as the {\em counting filter},
is known in the literature~\cite{GL89,JTU96,BYN02} 
and has been used for $k$-mismatches and $k$-differences.
The idea behind the filter is straightforward and is based upon 
the observation that (in our problem) if the pattern
$p$ has an approximate occurrence (possibly involving inversions and translocations)
starting at position $s$ of the text then the $m$-substring
of the text $t[s\ldots s+m-1]$ is a permutation of the pattern.

Then the \fgg algorithm identifies the set $\Gamma_{p,t}$ of all candidate 
positions $s$ in the text such that the substring $t[s\ldots s+m-1]$ is 
a permutation of the characters in $p$ and, for each $s\in \Gamma_{p,t}$, 
executes a verification procedure in order to check the approximate occurrence.

Before entering into details we need to introduce some additional notations.
Given two strings $w$ and $z$, we define a distance function $\delta(w,z)$ as
$$
  \delta(w,z) = \displaystyle \sum_{c\in \Sigma} \Abs{\occ_w(c) - \occ_z(c)}.
$$
Obviously, if $\slength{w} = \slength{z}$, then $\delta(w,z)=0$ iff $w$ is a permutation of $z$.

For each position $s$ in the text, with $0\leq s \leq n-m$, we define a function $G_s: \Sigma \rightarrow N$, as
$$
  G_s(c) =  \occ_p(c) - \occ_{t(s,m)}(c)
$$
for $c\in \Sigma$, and where we set $t(s,m) = t[s\ldots s+m-1]$.

Finally we define, for each position $s$, the distance value $\delta_s$ as follows
$$
  \delta_{s} = \delta(p,t_s) =  \displaystyle \sum_{c\in \Sigma} \Abs{\occ_p(c) - \occ_{t(s,m)}(c)} = \sum_{c\in \Sigma} \Abs{G_s(c)}.
$$
Then the set $\Gamma_{p,t}$ of all candidate positions in the text can be defined as
$$
  \Gamma_{p,t} = \displaystyle  \{ s\ \sht\ 0\leq s \leq n-m \textrm{ and } \delta_s = 0 \}.
$$

Observe that values $\delta_{s+1}$ and $\delta_{s}$ can differ only in the number of occurrences of characters $t[s]$ and $t[s+m]$.
Specifically we have $\occ_{t(s+1,m)}(t[s])\geq \occ_{t(s,m)}(t[s])-1$ and $\occ_{t(s+1,m)}(t[s+m])\leq \occ_{t(s,m)}(t[s+m])+1$.
Moreover, if $t[s]=t[s+m]$, the two functions $\occ_{t(s+1,m)}$ and $\occ_{t(s,m)}$ do not differ for any value.

Therefore, for each character $c\in \Sigma$, the value of $G_{s+1}(c)$ can be computed in constant time from $G_s(c)$ by using the following relation
$$
  G_{s+1}(c) = \left\{ \begin{array}{ll}
                G_s(c) -1	& \textrm{ if } c=t[s+m]\neq t[s]\\
                G_s(c) +1	& \textrm{ if } c=t[s] \neq t[s+m]\\
                G_s(c) 		& \textrm{ otherwise }\\
               \end{array}\right.
$$
which gives the following relation for computing $\delta_{s+1}$ from $\delta_{s}$ in constant time
$$
\begin{array}{ll}
    \delta_{s+1} = & \displaystyle \delta_{s} - \Abs{G_s(t[s])} - \Abs{G_s(t[s+m])} +\\ 
		  & + \Abs{G_{s+1}(t[s])} + \Abs{G_{s+1}(t[s+m])}.
\end{array}
$$

\begin{figure}[!t]
\begin{center}
\begin{scriptsize}
\begin{tabular}{|l|l|}
\hline
&\\
\setcounter{instr}{0}
\begin{tabular}{rl}
\multicolumn{2}{l}{\textsc{\fgg}$(p, m, t, n, \alpha, \beta)$}\\
\ninstr & \textbf{for} $c\in \Sigma$ \textbf{do} $\vect[c]\leftarrow 0$\\
\ninstr & \textbf{for} $s\leftarrow 0$ \textbf{to} $m-1$ \textbf{do}\\
\ninstr & \quad $\vect[p[s]]\leftarrow \vect[p[s]] + 1$\\
\ninstr & \quad $\vect[t[s]]\leftarrow \vect[t[s]] - 1$\\
\ninstr & $\delta\leftarrow 0$\\
\ninstr & \textbf{for} $c\in \Sigma$ \textbf{do} $\delta \leftarrow \delta+ \abs{\vect[c]}$\\
\ninstr & \textbf{for} $s\leftarrow 0$ \textbf{to} $n-m$ \textbf{do}\\
\ninstr & \quad \textbf{if} $\delta=0$ \textbf{then}\\
\ninstr & \quad \quad \textsc{Verify}$(p,m,t,s,\alpha,\beta)$\\
\ninstr & \quad $a\leftarrow t[s]$\\
\ninstr & \quad $b\leftarrow t[s+m]$\\
\ninstr & \quad $\delta\leftarrow \delta - \abs{\vect[a]} - \abs{\vect[b]}$\\
\ninstr & \quad $\vect[a] \leftarrow \vect[a] + 1$\\
\ninstr & \quad $\vect[b] \leftarrow \vect[b] - 1$\\
\ninstr & \quad $\delta\leftarrow \delta + \abs{\vect[a]} + \abs{\vect[b]}$\\
\ninstr & \textbf{if} $\delta=0$ \textbf{then}\\
\ninstr & \quad \textsc{Verify}$(p,m,t,n-m,\alpha,\beta)~~~~$\\
&\\
&\\
\end{tabular}

&
\setcounter{instr}{0}
\begin{tabular}{rl}
\multicolumn{2}{l}{\textsc{Verify}$(p, m, t, s, \alpha, \beta)$}\\
\ninstr & $\gamma = \min(\alpha, \beta)$\\
\ninstr & \textbf{for} $i \leftarrow 0$ \textbf{to} $m-1$ \textbf{do}\\
\ninstr & \quad \textbf{for} $j \leftarrow \max(0, i-\gamma)$ \textbf{to} $\min(m-1, i+\gamma)$ \textbf{do}\\
\ninstr & \quad \quad $F[i,j] \leftarrow I[i,m-j-1] \leftarrow 0$\\
\ninstr & \quad \quad \textbf{if} $(p[i]=t[s+j])$ \textbf{then}\\
\ninstr & \quad \quad \quad \textbf{if} $(i=0$ or $j=0)$ \textbf{then} $F[i,j]\leftarrow 1$\\
\ninstr & \quad \quad \quad \textbf{else} $F[i,j]\leftarrow F[i-1,j-1]+1$\\
\ninstr & \quad \quad \textbf{if} $(p[i]=t[s+m-j-1])$ \textbf{then}\\
\ninstr & \quad \quad \quad \textbf{if} $(i=0$ or $j=0)$ \textbf{then} $I[i,m-j-1]\leftarrow 1$~~~~\\
\ninstr & \quad \quad \quad \textbf{else} $I[i,m-j-1]\leftarrow I[i-1,m-j]+1$\\
\ninstr & \quad \textbf{if} $(p[i]=t[s+i]$ and $(i=0$ or $S[i-1]=1$))\\
\ninstr & \quad \textbf{then} $S[i] \leftarrow 1$ \textbf{else} $S[i]\leftarrow 0$\\
\ninstr & \quad \textbf{for} $k \leftarrow 1$ \textbf{to} $\min(\alpha,\lfloor \frac{i+1}{2}\rfloor)$ \textbf{do}\\
\ninstr & \quad \quad \textbf{if} $(F[i,i-k]\geq k$ and $F[i-k,i]\geq k)$ \textbf{then} \\
\ninstr & \quad \quad \quad \textbf{if} $(i<2k$ or $ S[i-2k]=1)$ \textbf{then} $S[i] \leftarrow 1$\\
\ninstr & \quad \textbf{for} $k \leftarrow 2$ \textbf{to} $\min(\beta, i+1)$ \textbf{do}\\
\ninstr & \quad \quad \textbf{if} $(I[i,i-k+1]\geq k)$ \textbf{then}\\
\ninstr & \quad \quad \quad \textbf{if} $(i<k$ or $S[i-k]=1)$ \textbf{then} $S[i] \leftarrow 1$\\
\ninstr & \textbf{if} $(S[m-1]=1)$ \textbf{then} Output($s$)\\

\end{tabular}\\

&\\
\hline
\end{tabular}
\end{scriptsize}
\caption{(on the left) The \fgg algorithm for the approximate string matching problem with inversions and translocations and
(on the right) the verification procedure.}
\label{fig:fgg}
\end{center}
\end{figure}

Fig.\ref{fig:fgg} shows the pseudocode of the \fgg algorithm (on the left) and the 
verification procedure (on the right).
Note that the main loop of \fgg has only one conditional and the integer {\em abs} 
function is translated by modern compilers (including \texttt{GNU C Compiler}) 
into branchless code.

The verification procedure is based on dynamic programming.
The algorithm uses two matrices, $F$ and $I$, both of size $m^2$, in order to compute occurrences of
factors and inverted factors of $p$, respectively, in the substring $t[s\ldots s+m-1]$.
More formally we define
$$
\begin{array}{ll}
F[i,j] & = \displaystyle \max\{k\ \sht\ p[i-k+1\ldots i] = t[s+j-k+1\ldots s+j]\}, \textrm{ and}\\
I[i,j] & = \displaystyle \max\{k\ \sht\ p[i-k+1\ldots i] = (t[s+j\ldots s+j+k-1])^r\}
\end{array}
$$
for $0\leq i< m$ and $\max(0, i-\gamma) \leq j \leq \min(m-1, i+\gamma)$, where $\gamma=\min(\alpha, \beta)$.
Moreover a vector $S$, of size $m$, is maintained in order to compute the $md$-matches of all prefixes of the pattern in $t[s\ldots s+m-1]$.
More formally, for $0\leq i <m$, we have
$S[i] = 1$  if $p_i \mdsuff t[s\ldots s+i]$ and $S[i]=0$ otherwise.

The following recursive relations are used for computing $F$ and $I$.
$$
\begin{array}{ll}
F[i,j] & = \left\{ \begin{array}{ll}
                0	& \textrm{ if } p[i] \neq t[s+j] \\
                F[i-1,j-1]+1 & \textrm{ if } i>0,j> \max(0, i-\alpha) \textrm{ and } p[i] = t[s+j]\\
                1	& \textrm{ otherwise }\\
               \end{array}\right. \\[0.5cm]
I[i,j] & = \left\{ \begin{array}{ll}
                0	& \textrm{ if } p[i] \neq t[s+j]\\
                I[i-1,j+1]+1	& \textrm{ if } i> 0, j<\min(m-1, i+\beta) \textrm{ and } p[i] = t[s+j]\\
                1	& \textrm{ otherwise }\\
               \end{array}\right.
\end{array}
$$

Finally the vector $S$ is computed, for increasing $i = 1\ldots m-1$ ($S[i]$ is set to 0)
according to the following (recursive) formula.
The value of $S[i]$ is set to $1$ iff one of the following conditions holds:
\begin{enumerate}[-]
 \item $p[i] = t[s + i]$ and $(i = 0$ or $S[i - 1] = 1)$;
 \item $F [i, i - k] \geq k$, $F [i - k, i] \geq k$ and $(i < 2k$ or $S[i - 2k] = 1)$, for $1\leq k \leq \min(\alpha,\lfloor \frac{i+1}{2}\rfloor)$;
 \item $I[i, i - k + 1] \geq k$ and $(i < k$ or $S[i - k] = 1)$, for $1\leq k \leq \min(\beta,i+1)$.
\end{enumerate}

Then $p$ has an $md$-match starting at position $s$ of the text
if $S[m-1]=1$ at the end of the verification procedure with parameter $p$, $t$ and $s$.

Observe that for computing the entry of position $i$ in $S$ only the last $\beta$ entries of the $(i-1)$th row of $I$ are needed, 
while only the last $\alpha$ entries of the $(i-1)$th row of $F$ and of the $(i-1)$th column of $F$ are needed. 
Similarly only the last $\max(2\alpha, \beta)$ entries of the vector $S$ are needed for computing the value $S[i]$.
Moreover, both for $I$ and $F$, the computation of the $i$th row (column)
needs only the values in the $(i-1)$th row (column) of the matrix.

It is thus straightforward to reduce the space requirements of the verification phase
to $\bigO(\max(\alpha, \beta))$. This is done by maintaining, for each iteration,
only two rows of $I$ and only two rows and two columns of $F$, each of size $\max(\alpha, \beta)$.

The verification time and space costs are thus 
$\bigO(m \max(\alpha, \beta))$ and $\bigO(\max(\alpha, \beta))$, respectively,
leading to overall $\bigO(nm\max(\alpha, \beta))$ worst case time complexity and 
$\bigO(\max(\alpha, \beta, \sigma))$ space complexity for the \fgg algorithm.

\section{Average Case Time Analysis}\label{sec:average}
Next, we evaluate the average time complexity of the \fgg algorithm.
In our analysis we assume the uniform distribution and independence of characters.
We first assume that $m = \omega(\sigma^{\bigO(1)})$,
Then we prove the more simple case when $m \leq \sigma$.

Our verification procedure takes $\bigO(m^2)$ (worst-case) time per location.
To obtain linear average time, we must thus bound the probability of 
having permuted subsequences of length $m$ with $\bigO(1/m^2)$.
We will find conditions upon which this happens.\footnote{The paper~\cite{BYN02} 
contains an analysis of the counting filter, 
in the $k$-differences problem. 
Unfortunately, the analysis seems to be flawed, which was admitted 
in discussion by the second author of the cited paper (G. Navarro).}

Suppose $m = \omega(\sigma^{\bigO(1)})$, we define $k = m/\sigma$ and, without loss of generality, we assume that $\sigma$ divides $m$.
For each text position $s$, with $0\leq s\leq n-m$, the probability 
that the $m$-substring of the text, beginning at position $s$, is a permutation of the pattern $p$ is exactly
\begin{equation} \label{prob1}
    \Pr\{s \in \Gamma_{p,t}\} = \frac{      {m \choose occ(c_0)}
                {m-occ(c_0) \choose occ(c_1)}
                {m-occ(c_0)-occ(c_1) \choose occ(c_2)}
                \ldots
                {occ(c_{\sigma-1}) \choose occ(c_{\sigma-1})} 
         } 
         {\sigma^m}.
\end{equation}

Now, it is easy to notice that the probability given in (\ref{prob1}) is maximized when $occ(c_i) = k$ 
for all $i$.
We can thus write:
$$
  \Pr\{s \in \Gamma_{p,t}\} \leq 
  \frac{ {m \choose k}{m-k \choose k}{m-2k \choose k}\ldots{k \choose k} } {\sigma^m} =
  \frac{m!}{(k!)^\sigma \sigma^m}.
$$


We make use of Stirling's approximation for both $m!$ and $k!$, recalling that 
$k = m/\sigma$:
$$
  \frac{m!}{(k!)^\sigma \sigma^m} = \Theta\left(\frac{\sqrt{2\pi m}(m/e)^m}
                                       {(\sqrt{2\pi (m/\sigma)}(m/(e\sigma))^{m/\sigma})^\sigma \sigma^m} \right) =
  \Theta\left(\frac{\sqrt{2\pi m}}{\left(\sqrt{2\pi (m/\sigma)}\right)^\sigma} \right).
$$

Let us upper-bound $\sqrt{2\pi} / (\sqrt{2\pi})^\sigma$ with 1 and remove it.
We have:
$$
  \Theta\left( \frac{\sqrt{m}}{\left(\sqrt{m/\sigma}\right)^\sigma} \right) = \Theta\left(\frac{\sigma^{\sigma/2}}{m^{(\sigma-1)/2}}\right).
$$

Let us assume $m \geq \sigma^4$ (we recall that $m = \omega(\sigma^{\bigO(1)})$).
Then 
$\sigma^{\sigma/2}/m^{(\sigma-1)/2}$ is less than or equal to $1/\sigma^{1.5\sigma-2}$.
Note that if we take a larger lower bound on $m$, e.g., $\sigma^8$, then our upper bound 
gets even smaller, namely $1 / \sigma^{3.5\sigma-4}$ in this example.
All in all, we have 
$$
  \Pr\{s \in \Gamma_{p,t}\} = \bigO(1 / \sigma^{\bigO(\sigma)}) = \bigO(1/m^2) 
$$
for any $\sigma = \Omega(\log m / \log\log^{1-\varepsilon} m)$, where $\varepsilon > 0$. 

Suppose now that $m \leq \sigma$.\footnote{Note that for the more general case of 
$m = \sigma^{\bigO(1)}$ there exists already an average-case linear algorithm~\cite{PSC2010-4}, 
so this part of the analysis is only to find properties of the currently presented 
algorithm.}
Then the probability
that the $m$-substring of the text, beginning at position $s$, is a permutation of the pattern $p$ is
$$
  \Pr\{s \in \Gamma_{p,t}\} \leq \frac{m!}{\sigma^m} \leq \frac{m!}{m^m}
$$

If we make use again of Stirling's approximation for $m!$, we obtain
$$
  \Pr\{s \in \Gamma_{p,t}\} < \sqrt{2\pi} \frac{m^{m+1}}{e^m m^m} = \sqrt{2\pi} \frac{m}{e^m} = \bigO(1/m^2).
$$

Thus the overall average time complexity of the \fgg algorithm, assuming $\sigma = \Omega(\log m / \log\log^{1-\varepsilon} m)$, 
is given by the following relation:
\begin{eqnarray*}
  T(n,m,\sigma) & = & \bigO(\sigma + m) + \sum_{s=0}^{n-m} \Pr\{s \in \Gamma_{p,t}\} \cdot \bigO(m^2) \\
                & = & \bigO(\sigma + m) + (n-m+1) \cdot \bigO(1/m^2) \cdot \bigO(m^2) = \bigO(n).
\end{eqnarray*}

\section{Experimental results}\label{sec:exp}
In this section we evaluate the performance of the following algorithms:
\begin{itemize}
 \item The \msampling algorithm~\cite{PSC2010-4} (\ms)
 \item The \fgg algorithm using the \msampling algorithm for verification (\fgga)
 \item The \fgg algorithm as shown in Fig.\ref{fig:fgg} (\fggb)
\end{itemize}
All algorithms have been implemented in \textsf{C} and 
compiled with the \texttt{GNU C Compiler 4.2},
using the options \texttt{-O2 -fno-guess-branch-probability}.  
All tests have been performed on a 2
GHz Intel Core 2 Duo and running times have been measured with a
hardware cycle counter, available on modern CPUs.  We used the
following input files:
\begin{enumerate}[(i)]
    \item four random texts of $2,000,000$ characters with a uniform distribution over alphabets of dimension $\sigma$,
    with $\sigma\in\{4,8,16,32\}$ respectively,
    \item a protein sequence of $2,900,352$ characters from the \textit{Saccharomyces
    cerevisiae} genome (with $\sigma=20$),\footnote{
    \url{http://data-compression.info/Corpora/ProteinCorpus/}}
    \item a genome sequence of $4,638,690$ base pairs of
    \textit{Escherichia coli} (with $\sigma=4$).\footnote{
    \url{http://corpus.canterbury.ac.nz/}}
\end{enumerate}
For each input file, we have
generated seven sets of $200$ patterns of fixed length $m$ randomly
extracted from the text (with at least one occurrence in the text), 
for $m$ ranging over the values $8$, $16$,
$32$, $64$, $128$, $256$, $512$. 
For each set of patterns we reported the mean time over $200$ runs,
expressed in milliseconds.

\begin{center}

\begin{scriptsize}
\begin{tabular*}{0.31\textwidth}{@{\extracolsep{\fill}}|l|cccc|}
\hline
\multicolumn{5}{|l|}{\textsf{Random text with $\sigma=4$}}\\
\hline
$m$ & \ms & \fgga & \fggb &\\
\hline
8 & 254.78 & 48.53 & 73.73 &\\
16 & 350.25 & 50.05 & 103.09 &\\
32 & 441.05 & 44.20 & 102.04 &\\
64 & 528.35 & 43.83 & 140.18 &\\
128 & 645.36 & 43.20 & 208.05 &\\
256 & 868.13 & 41.84 & 273.47 &\\
512 & 1273.13 & 44.71 & 349.57 &\\
\hline
\end{tabular*}~~~
\begin{tabular*}{0.31\textwidth}{@{\extracolsep{\fill}}|l|cccc|}
\hline
\multicolumn{5}{|l|}{\textsf{Random text with $\sigma=8$}}\\
\hline
$m$ & \ms & \fgga & \fggb &\\
\hline
8 & 155.39 & 29.57 & 29.78 &\\
16 & 193.91 & 29.21 & 28.98 &\\
32 & 241.54 & 29.20 & 28.72 &\\
64 & 309.26 & 29.33 & 28.75 &\\
128 & 377.17 & 29.68 & 29.16 &\\
256 & 525.96 & 30.75 & 30.89 &\\
512 & 770.45 & 34.14 & 37.73 &\\
\hline
\end{tabular*}~~~
\begin{tabular*}{0.31\textwidth}{@{\extracolsep{\fill}}|l|cccc|}
\hline
\multicolumn{5}{|l|}{\textsf{Random text with $\sigma=16$}}\\
\hline
$m$ & \ms & \fgga & \fggb &\\
\hline
8 & 115.27 & 28.45 & 28.55 &\\
16 & 137.27 & 28.48 & 28.54 &\\
32 & 161.25 & 28.51 & 28.57 &\\
64 & 211.75 & 28.65 & 28.66 &\\
128 & 273.53 & 28.94 & 29.01 &\\
256 & 371.65 & 29.86 & 30.34 &\\
512 & 536.40 & 32.85 & 35.79 &\\
\hline
\end{tabular*}\\[0.2cm]
\end{scriptsize}

\begin{scriptsize}
\begin{tabular*}{0.31\textwidth}{@{\extracolsep{\fill}}|l|cccc|}
\hline
\multicolumn{5}{|l|}{\textsf{Random text with $\sigma=32$}}\\
\hline
$m$ & \ms & \fgga & \fggb &\\
\hline
8 & 93.80 & 28.18 & 28.52 &\\
16 & 110.64 & 28.20 & 28.53 &\\
32 & 128.80 & 28.25 & 28.55 &\\
64 & 169.25 & 28.42 & 28.61 &\\
128 & 197.24 & 28.65 & 28.93 &\\
256 & 259.77 & 29.45 & 30.23 &\\
512 & 398.20 & 32.07 & 35.11 &\\
\hline
\end{tabular*}~~~
\begin{tabular*}{0.31\textwidth}{@{\extracolsep{\fill}}|l|cccc|}
\hline
\multicolumn{5}{|l|}{\textsf{Escherichia coli}}\\
\hline
$m$ & \ms & \fgga & \fggb &\\
\hline
8 & 593.49 & 117.79 & 184.48 &\\
16 & 781.76 & 108.53 & 208.50 &\\
32 & 976.79 & 99.88 & 222.19 &\\
64 & 1188.58 & 94.64 & 267.01 &\\
128 & 1484.03 & 84.16 & 252.17 &\\
256 & 2005.00 & 80.40 & 257.70 &\\
512 & 2929.90 & 83.36 & 299.49 &\\
\hline
\end{tabular*}~~~
\begin{tabular*}{0.31\textwidth}{@{\extracolsep{\fill}}|l|cccc|}
\hline
\multicolumn{5}{|l|}{\textsf{Saccharomyces cerevisiae}}\\
\hline
$m$ & \ms & \fgga & \fggb &\\
\hline
8 & 163.25 & 41.38 & 41.45 &\\
16 & 192.64 & 41.39 & 41.45 &\\
32 & 224.27 & 41.44 & 41.48 &\\
64 & 297.01 & 41.56 & 41.60 &\\
128 & 376.27 & 41.88 & 41.91 &\\
256 & 506.88 & 42.79 & 43.25 &\\
512 & 738.19 & 45.72 & 48.65 &\\
\hline
\end{tabular*}\\[0.2cm]
\end{scriptsize}

\end{center}

The experimental results show that the filtering strategy is quite
effective and allows to dramatically 
speed up, by a factor of at most $30$,
the computation of the $md$-matches of a given pattern. It is worth
observing that for very small alphabets the \fgga algorithm, based
on $\msampling$, is faster than the \fggb algorithm, based on the 
dynamic programming verification, while in the other cases 
the two algorithms have almost the same speed. 
In the following tables we report the mean, over the $200$ runs, of the
number of pattern's permutations found per text position.

\begin{center}

\begin{scriptsize}
\begin{tabular*}{0.30\textwidth}{@{\extracolsep{\fill}}|lc|}
\hline
\multicolumn{2}{|l|}{\textsf{Random text ($\sigma=4$)}}\\
\hline
$m$ & \# candidate \\
\hline
8 & 0.013621 \\
16 & 0.006399 \\
32 & 0.001837 \\
64 & 0.000720 \\
128 & 0.000285 \\
256 & 0.000093 \\
512 & 0.000029 \\
\hline
\end{tabular*}~~~
\begin{tabular*}{0.30\textwidth}{@{\extracolsep{\fill}}|lc|}
\hline
\multicolumn{2}{|l|}{\textsf{Random text ($\sigma=8$)}}\\
\hline
$m$ & \# candidate \\
\hline
8 & 0.000410 \\
16 & 0.000037 \\
32 & 0.000004 \\
64 & 0.000001 \\
128 & 0.000001 \\
256 & 0.000001 \\
512 & 0.000001 \\
\hline
\end{tabular*}~~~
\begin{tabular*}{0.30\textwidth}{@{\extracolsep{\fill}}|lc|}
\hline
\multicolumn{2}{|l|}{\textsf{Random text ($\sigma=16$)}}\\
\hline
$m$ & \# candidate \\
\hline
8 & 0.000004 \\
16 & 0.000001 \\
32 & 0.000001 \\
64 & 0.000001 \\
128 & 0.000001 \\
256 & 0.000001 \\
512 & 0.000001 \\
\hline
\end{tabular*}\\[0.2cm]
\end{scriptsize}
Average number of candidate positions for each text character on random texts with $\sigma=4$ (on the left) $\sigma=8$ (in the center) $\sigma=16$ (on the right)\\[.3cm]
\end{center}

Observe that, while for small alphabets
the number is non negligible also for long patterns, for large enough
alphabets it is always insignificant.

\section{Acknowledgement}\label{sec:ack}
The work was partially supported (the first author) 
by the Polish Ministry of Science and Higher Education 
under the project N N516 441938.

\begin{small}
\bibliographystyle{abbrv}
\bibliography{IPLgfg}
\end{small}

\end{document}